%% file: Manuscript.tex
\documentclass[%
 aip,
 amsmath,amssymb,
reprint,%
]{revtex4-1}

\input{definitions}

\usepackage{graphicx}
\usepackage{dcolumn}
\usepackage{bm}

\usepackage[utf8]{inputenc}
\usepackage[T1]{fontenc}
\usepackage{mathptmx}
\usepackage{siunitx}
\usepackage{xcolor}
\DeclareSIUnit\gauss{G}

\begin{document}

\preprint{AIP/123-QED}

\title{Semi-Implicit Continuum Kinetic Modeling of Weakly Collisional Parallel Transport in a Magnetic Mirror}

\author{M. Dorf}
\email{dorf1@llnl.gov}
\author{M. Dorr}
\author{V. Geyko}
\author{D. Ghosh}
\author{M. Umansky}
\author{J. Angus}
\affiliation{ 
Lawrence Livermore National Laboratory, Livermore, California 94550 USA 
}

\date{\today}

\begin{abstract}

We present implicit-explicit (IMEX) kinetic simulations of weakly collisional parallel plasma transport in magnetic mirror configurations using the continuum code \textsc{COGENT}. The numerical scheme employs a Jacobian-free Newton--Krylov method with algebraic multigrid preconditioning to overcome the severe time-step limitations imposed by strong mirror forces in fully explicit schemes. Applied to parameters relevant to the WHAM mirror experiment, the IMEX approach enables time steps up to $2.5 \times 10^4$ times larger than those permitted by explicit methods, resulting in a 2500x speedup in 1D--2V simulations of parallel transport with kinetic ions and Boltzmann electrons. Additionally, a reduced bounce-averaged model for a square mirror is implemented to support the computationally intensive fully kinetic simulations. The bounce-averaged formulation is used to evaluate the numerical convergence of the velocity-space discretization algorithms and to assess the role of the collision model by comparing simulations employing the nonlinear Fokker--Planck and the simplified Lenard--Bernstein--Dougherty collision operators.

\end{abstract}

\maketitle

\section{\label{sec:level1} Introduction}

In the past several decades, magnetic fusion research has been heavily focused on large-scale toroidal confinement devices like tokamaks~\cite{wesson:2011} and stellarators~\cite{warmer:2024}. However, recent technological advances including the development of high-temperature superconducting (HTS) magnets appropriate for fusion applications\cite{whyte:2019}, together with improvements in the theoretical understanding of magnetized plasma stability and transport, have enabled researchers to explore and experiment with re-emerging alternative fusion concepts such as magnetic mirrors \cite{endrizzi:2023, forest:2024, ivanov:2017, ball:2025}. Being more compact and thereby less expensive, magnetic mirrors attract particular attention from academia and private investors. A notable example of a modern mirror experiment is the Wisconsin HTS Axisymmetric Mirror\cite{endrizzi:2023} (WHAM), which explores the benefits of strong magnetic fields with high mirror ratio for the classical weakly collisional confinement approach.

The success of alternative magnetized fusion concepts, such as magnetic mirrors, relies heavily on advanced modeling capabilities that can provide scientific insights and assess scalability to reactor parameters. This motivates the adaptation of advanced gyrokinetic computational tools, which have been extensively developed over the past several decades for the mainstream tokamak approach, to mirror applications. Although modeling certain mirror-specific high-frequency processes, e.g., the drift-cyclotron loss-cone instability, requires resolving cyclotron ion motion \cite{tran:2025}, the analysis of lower-frequency processes with $\omega \ll \omega_{ci}$, such as collisional transport and gradient-driven drift modes, can be facilitated by employing the gyro-averaging formalism. Here, $\omega_{ci}$ denotes the ion cyclotron frequency. A recent study by M. Francisquez \textit {et al} \cite{francisquez:2023} outlines the challenges faced by continuum gyrokinetic codes in modeling HTS mirrors and demonstrates the application of the gyrokinetic code Gkeyll to simulate 1D-2V parallel transport in mirror geometry for parameters characteristic of the WHAM experiment\cite{endrizzi:2023}. We also note 2D–2V gyrokinetic simulations\cite{wang:2024}, performed with the particle-in-cell GTC-X code, that investigate the penetration of divertor electrostatic biasing in the scrape-off layer of a field-reversed configuration, which shares linear-geometry features with magnetic mirrors.

Although the presence of an open-field line region in a tokamak edge implies a certain level of geometrical similarity with a magnetic mirror, the physics parameters can nevertheless be quite different. Classical mirror plasmas are significantly less collisional than the edge of a tokamak. For instance, for the parameters characteristic of a mid-size tokamak (e.g., DIII-D) edge~\cite{chan:2000}, $n_i \sim \SI{e19}{\per\cubic\meter}$, $T_i \sim \SI{100}{eV}$, $L_\parallel \sim \SI{6}{m}$, $m_i = 2 m_p$, the Knudson number, measuring the ratio of the ion-ion mean free path, $\lambda_{ii}$, to a length scale in variation of background plasma profiles along the magnetic field, $L_\parallel$, is given by $K_{DIII-D} \sim 3$. Here, $n_i$ is the ion density, $T_i$ is the ion temperature, and $m_i$ and $m_p$ denote the ion and proton mass, respectively. In contrast, the WHAM mirror parameters,  $n_i \sim \SI{3e19}{\per\cubic\meter}$, $T_i \sim \SI{8}{\kilo\electronvolt}$, $L_\parallel \sim \SI{2}{m}$, $m_i = 2m_p$, correspond to $K_{WHAM} \sim 1.5\times10^4$. This means that the ion transit time is four orders of magnitude shorter than the ion-ion collision time, thereby making the problem of collisional mirror transport intrinsically multiscale in time integration. It is also important to note that the presence of loss-cone regions and high-energy beams in the velocity phase space requires the use of \textit{full-F} computational methods that include background plasma evolution. 

The temporally multiscale nature of a weakly collisional mirror transport can be effectively handled by making use of the bounce-average formulation that averages over the fast ion transit time\cite{endrizzi:2023, forest:2024, petrov:2023}. Although this approach can significantly facilitate numerical simulations of plasma dynamics inside the mirror, it lacks the ability to efficiently describe the region outside the trap and the coupling of the confined mirror region to plasma-facing components. Moreover, the standard bounce-averaged formulation assumes a single trapped region---an assumption that is violated in tandem magnetic mirrors~\cite{dimov:2005, ryutov:1988} and even in simple mirrors with sloshing ion distributions~\cite{ryutov:2011}, where electrons can become trapped in local potential peaks near ion beam turning points. Therefore, first-principles gyrokinetic simulations are required for high-fidelity, integrated modeling.

It is interesting to note that the ion transit time, $\tau_{\parallel,i} \sim \SI{50}{\micro\second}$, and the electron transit time, $\tau_{\parallel,e} \sim \SI{1}{\micro\second}$, at the tokamak edge are both much shorter than the characteristic time of the anomalous radial plasma transport, $\tau_\text{transp} \sim \SI{1}{\milli \second}$. Furthermore, full-$F$ simulations are required to describe substantial deviations of a plasma distribution function from a local Maxwellian. Nevertheless, modern HPC gyrokinetic codes are capable of performing millisecond full-$F$ simulations of 2D-2V and even 3D-2V plasma transport using straightforward explicit time integration schemes~\cite{michels:2022}. In contrast to tokamaks, mirrors are distinguished by the presence of a strong mirror force, ${\bm F}_{\nabla B} = - \mu \nabla B$, which can impose a stringent CFL limit on the size of a stable time step for explicit time integration: $\Delta t_{\nabla B}^{\text{CFL}} < {m_i \Delta \vpll}/({\mu \nabla B})$. Here, $\Delta \vpll$ is the cell size in the parallel velocity direction, $\mu = m_i \vperp^2 /(2B)$ is the magnetic moment, and for a typical WHAM simulation, $\Delta t_{\nabla B} \sim 0.1\,\text{ns}$. As a result, explicit time integration of a weakly collisional mirror plasma on the ion-ion collisional timescale (tens of milliseconds) becomes prohibitively expensive. Recent explicit numerical simulations of 1D-2V collisional transport performed with the Gkeyll code required approximately 74 hours of wall time on 288 cores to simulate $\SI{72}{\micro\s}$ of plasma dynamics involving kinetic ions and Boltzmann electrons~\cite{francisquez:2023}. Simulations with kinetic electrons, which are also considered in that work, are significantly more expensive due to stiff electron time scales. Although a substantial, 20x, speedup was demonstrated by introducing \textit{ad hoc} softening of the $\nabla B$ force, explicit simulations spanning tens of milliseconds remain prohibitively costly for explicit methods.

Here, the feasibility of multiscale collisional mirror transport simulations is explored by making use of the implicit-explicit (IMEX) time integration framework implemented in the continuum gyrokinetic code COGENT, originally developed for tokamak edge modeling~\cite{dorf:2021}. The IMEX time integration algorithm used in COGENT~\cite{ghosh:2018} is based on semi-implicit additive Runge–Kutta (ARK) methods~\cite{kennedy:2003} and can provide consistent high-order time integration, including implicit treatment of selected stiff terms. It employs the Jacobian-free Newton-Krylov (JFNK) approach~\cite{knoll:2004} to handle nonlinearities and utilizes preconditioning to improve convergence properties. Declaring the gyrokinetic Vlasov term as implicit enables stable time integration with stepping over the stiff time scales including advection, i.e., acceleration, in the parallel velocity direction due to strong mirror force and the parallel streaming. On the other hand, efficient preconditioning of the Vlasov operator (even in a simple case with a fixed advection velocity) requires solving a linear system in the full phase space dimension with a non-symmetric and indefinite coefficient matrix. COGENT solves this linear system using the Approximate Ideal Restriction (AIR) option in the BoomerAMG algebraic multigrid solver contained in the Hypre linear solver library~\cite{Hypre}. The use of multigrid methods to solve nonsymmetric indefinite systems has historically been highly problematic, but the recent development of the AIR approach~\cite{manteuffel:2018, manteuffel:2019}, including several variants, provides a way to extend the benefits of multigrid algorithms beyond the symmetric, positive-definite systems for which they are more commonly used. To further enhance the efficiency of the implicit Vlasov solver, COGENT allows the preconditioner to be defined using a lower-order discretization (first-order upwind, UW1) than that used for the Vlasov operator itself (fifth-order upwind, UW5). The low-order preconditioner yields a sparser matrix, for which robust AMG solver performance is observed, while also providing good efficiency even for relatively large time steps. 

The IMEX COGENT time integration scheme is applied to 1D-2V simulations of parallel collisional transport using parameters characteristic of the WHAM experiment. The simulations employ the kinetic Vlasov equation for the ion species coupled with the nonlinear ion-ion Fokker-Planck collisions and a high-energy beam source. A Boltzmann electron model is used to describe self-consistent electrostatic potential variations. The implicit simulations enable time steps up to $2.5 \times 10^4$ larger than the CFL-limited time step and demonstrate speedups of about 2500x and 1800x compared to the corresponding explicit COGENT simulations and previously reported explicit Gkeyll simulations with Boltzmann electrons~\cite{francisquez:2023}, respectively. Additionally, a fully implicit, simplified bounce-averaged (0D-2V) model for a square mirror (the BASM model) is implemented in COGENT to provide insights into the more computationally expensive 1D-2V simulations. The BASM model is verified by reproducing analytic results for the Pastukhov problem~\cite{pastukhov:1974, cohen:1978} and is used to examine the numerical convergence of the velocity-space discretization algorithms, which are challenged by the presence of steep gradients at the loss-cone boundary for a weakly collisional mirror. The BASM model is also used to assess the role of the ion-ion collision model by comparing simulations using the Fokker-Planck operator~\cite{dorf:2014} with those using the Lenard–Bernstein–Dougherty (LBD) model~\cite{francisquez:2022, ulbl:2022, angus:2012}, which is often employed in full-F gyrokinetic modeling due to its simplicity. 

The paper is organized as follows: the results of numerical simulations obtained using the simplified bounce-averaged model and the fully kinetic model are presented in Secs.~II and III, respectively, and Sec.~IV provides the conclusions.
 
\section{Bounce-averaged (0D-2V) simulations}

\subsection{Simulation model}

\begin{figure*}
\includegraphics[width=0.9\textwidth]{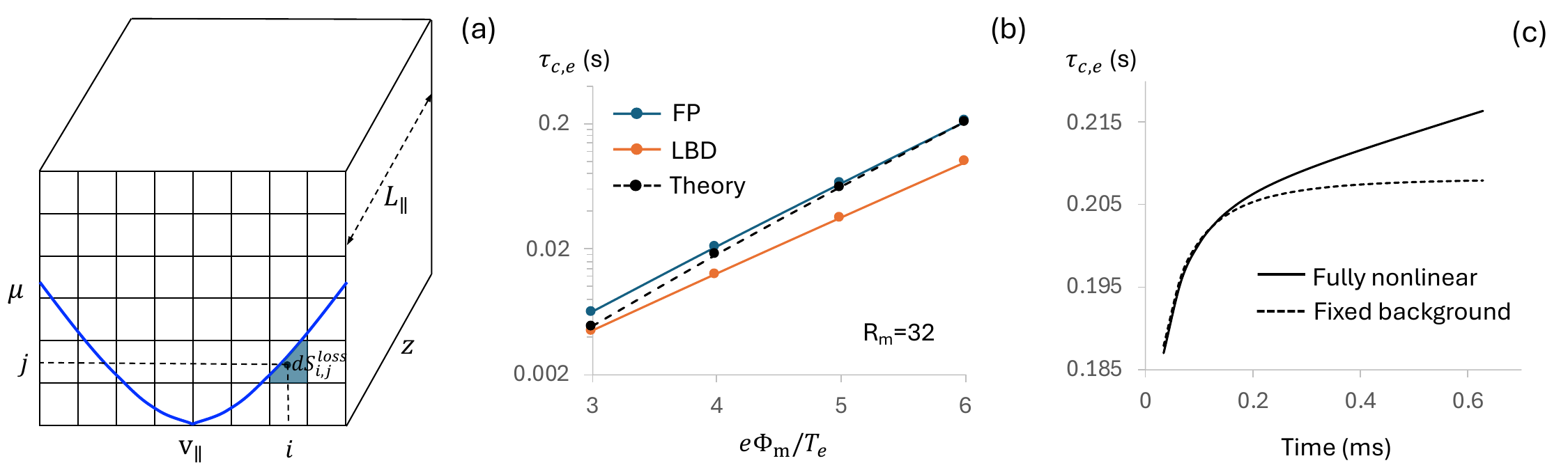}
\caption{Bounce-averaged (0D-2V) simulations. (a) Schematic of the bounce-averaged square-mirror (BASM) model. The blue curve illustrates the loss-cone boundary for $\Phi_m=0$. For the case where the loss-cone boundary intersects a cell $(i,j)$, the shaded area, $dS^{loss}_{i, j}$, is used in Eq.~(\ref{SinkTermDiscr}). (b) Collisional losses of the electrostatically-confined electron species for $R_m = 32$, $n=\SI{e19}{\m^{-3}}$, and $T_e = \SI{940}{\eV}$. The simulation results for the electron confinement time, $\tau_{c,e} = n_e |dn_e/dt |^{-1}$, obtained using the FP (blue) and LBD (orange) e-e collision models are compared with the analytical prediction in Eq.~(\ref{LossesAnalytic}). (c) Time history of the electron confinement time obtained using the full nonlinear FP model and the fixed-background FP model for $e\Phi_m/T_e =6$.}
 \label{Fig_BASM_schematic}
\end{figure*}

The bounce-averaged square-mirror (BASM) model describes a weakly collisional plasma confined in a square magnetic mirror. Under these assumptions, the kinetic distribution function for a plasma species can be considered uniform along the magnetic field, $f = f(\vpll, \mu)$, and the governing kinetic equation takes the following form:
\begin{equation}
\frac{\partial f}{\partial t} = C[f] + \Gamma_{src} \left[ f \right] - \Gamma_{sink} \left[ f \right] .
\label{BASMeq}
\end{equation}
Here, $C[f]$ is the collision operator, $\Gamma_{src}$ represents a source term, e.g., high-energy beams, and 
\begin{equation}
\Gamma_{sink} [f] = (\vpll / L_\parallel) f ,
\label{SinkTerm}
\end{equation}
describes prompt parallel losses from the loss-cone region, e.g., due to collisional scattering. Here, $m$, $\vpll$ and $\mu = m \vperp^2 /(2B)$ denote the particle mass, parallel velocity, and magnetic moment, respectively; $L_\parallel$ is the trap length, and the weakly-collisional assumption implies that $\lambda \gg L_\parallel$, where  $\lambda$ is the collisional mean free path. The sink term in Eq.~(\ref{SinkTerm}) is applied to the velocity-space region outside the loss-cone boundary given by
\begin{equation}
m \vpll^2 /2 \ge \mu B_0 \left( R_m -1 \right) + q \Phi_m .
\label{LossConeBnd}
\end{equation}
Here, $q$ is the particle charge, $B_0$ is the magnetic field at the mirror center, $R_m$ is the mirror ratio, defined as the ratio of the maximum magnetic field (at the mirror throat) to the minimum field (at the mirror center), and $\Phi_m$ is the confining electrostatic potential. The boundary in Eq.~(\ref{LossConeBnd}) cuts through the cells of the $(\vpll,\mu)$ velocity grid (see Fig.~\ref{Fig_BASM_schematic}), and the following numerical implementation of the sink term is adopted 
\begin{equation}
\Gamma^{sink}_{i, j}  = \frac{ \mathrm{v}_{\parallel,i} } { L_\parallel } \times f_{i, j}  \times \frac{dS^{loss}_{i, j} }{d\vpll d\mu},
\label{SinkTermDiscr}
\end{equation}
where $dS^{loss}_{i, j}$ denotes the area of a phase-space cell in the loss-cone region. For the case where the loss-cone boundary intersects a cell $(i,j)$, a second-order accurate approximation is used to compute $dS^{loss}_{i, j}$ [see Fig.~\ref{Fig_BASM_schematic}(a)]. For all other cells, $dS^{loss}_{i, j} = d\vpll d\mu$.

The collision models considered in this work for like-species collisions include the nonlinear Fokker-Planck (FP) model described elsewhere~\cite{dorf:2014} and the reduced Lenard–Bernstein–Dougherty (LBD) model~\cite{angus:2012}, which is often employed in full-F gyrokinetic modeling due to its simplicity
\begin{eqnarray}
C_{LBD} \left[ f \right] \nonumber 
 &&= \nu \frac{\partial}{\partial \vpll} \left[ \left( \vpll - U_\parallel \right) f + \frac{T}{m} \frac{\partial f}{\partial \vpll}\right] \nonumber\\
 &&+  \nu \frac{\partial}{\partial \mu} \left[ 2 \mu f + 2 \frac{T}{B} \mu \frac{\partial f}{\partial \mu} \right]. 
  \label{LBDOp}
\end{eqnarray}
Here, the particle number density ($n$), parallel flow velocity ($U_\parallel$), and the temperature ($T$) moments are defined as 
\begin{equation}
n =  \frac {2 \pi} {m} \int f B d\vpll d{\mu}.
  \label{Nmom}
\end{equation}
\begin{equation}
U_\parallel =  \frac {2 \pi} {n m} \int \vpll f B d\vpll d{\mu} ,
  \label{UparMom}
\end{equation}
\begin{equation}
T=  \frac {2 \pi} {3 n} \int \left[ \left( \vpll-U_\parallel \right)^2 +\frac{2\mu B}{m} \right] f Bd\vpll d{\mu}.
  \label{Tmom}
\end{equation}
The collision frequency in Eq.~(\ref{LBDOp}) is given by 
\begin{equation}
\nu =  \frac {4 \pi^{1/2} n Z^4 e^4 } {3 m^{1/2} T^{3/2}}  \ln \Lambda.
  \label{nuLBD}
\end{equation}
such that the LBD collision model matches the temperature relaxation rate of the Boltzmann operator~\cite{ulbl:2022}. Here,  $Z$ is the species charge state and $e$ is the elementary charge. The Coulomb logarithm used for the LBD and FP models is defined in the present work as  
\begin{equation}
\ln \Lambda = 23 - \ln \left( 2^{1/2} Z^3 n_0^{1/2}T_0^{-3/2} \right),
  \label{LnLambda}
\end{equation}
where $n_0 [\text {cm}^{-3}]$ and $T_0 [\text{eV}]$ are the density and temperature normalization factors used in the simulations.

COGENT employs an implicit-explicit (IMEX) time integration algorithm~\cite{ghosh:2018} based on semi-implicit additive Runge–Kutta (ARK) methods~\cite{kennedy:2003}, which enables consistent high-order time integration and allows for the implicit treatment of selected stiff terms. Nonlinearities are addressed using the Jacobian-free Newton-Krylov (JFNK) approach~\cite{knoll:2004}, with preconditioning applied to enhance convergence properties. The preconditioning of the stiff collision terms is developed assuming that the Rosenbluth potentials for the FP operator and the distribution function moments (i.e., $n$, $U_\parallel$, $T$) in the LBD model [see Eq.~(\ref{LBDOp})] remain fixed during a time step update. Under these assumptions, the collisional operator reduces to a linear advection-diffusion operator, for which a sparse matrix is assembled, and the Gauss-Seidel method is used to solve the resulting linear system. To improve consistency between the preconditioner and the full collision operator, thereby enhancing overall simulation performance, the same "fixed-moment" assumption is applied when evaluating the actual collision term in Eq.~(\ref{BASMeq}). Namely, although a second-order time integration scheme is employed, the moments of the distribution function are updated only once at the end of each time step. While such an approximation reduces the formal time integration accuracy, these effects might be minimal given that the simulations target long transport timescales. Finally, a second-order discretization in velocity space is employed for the FP and LBD collision operators. 

\subsection{Verification: collisional mirror losses}

The BASM model is verified by analyzing collisional losses of electrostatically confined species in a square magnetic mirror for a weakly collisional case (see Fig.~\ref{Fig_BASM_schematic}). Here, an electron species is considered due to the relative simplicity of electron-ion collisions, which can be described using the Lorentz collision model. Electron-electron collisions are modeled with the full nonlinear Fokker-Planck operator. Good agreement [see Fig.~\ref{Fig_BASM_schematic}(b)] is obtained with the approximate analytic solution in a weakly collisional regime with $R_m \gg 1$ and $\Phi_m / eT_e \gg 1$, originally derived by Pastukhov~\cite{pastukhov:1974} and later corrected by Cohen \cite{cohen:1978}
\begin{equation}
\frac{dn_e}{dt} = - \frac{4}{\sqrt{\pi}} \frac{n_e \nu_e}{G\left(R_m\right)} \frac{T_e}{e \Phi_m} \exp \left( -\frac{e\Phi_m}{T_e} \right) I\left( \frac{T_e}{e\Phi_m} \right),
  \label{LossesAnalytic}
\end{equation}
where $I(x)$ is related to the error function
\begin{equation}
I\left(x\right) = 1 + \frac{1}{2} \left( \pi x \right)^{1/2} e^{1/x} \left[ 1- \text{erf} \left( x^{-1/2}\right)\right],
  \label{I(x)}
\end{equation}
and where 
\begin{equation}
G\left(R_m \right) \approx \frac{2R_m+1}{2R_m} \ln \left( 4R_m+2 \right), 
  \label{G(R)}
\end{equation}
for $R_m \gg 1$. The electron collision frequency in Eq.~(\ref{LossesAnalytic}) is given by 
\begin{equation}
\nu_e =  \frac {\sqrt{2} \pi n e^4 } {m_e^{1/2} T_e^{3/2}}  \ln \Lambda.
  \label{nuCohen}
\end{equation}
For comparison, the electron confinement time, $\tau_{c,e} = n_e |dn_e/dt |^{-1}$, is also computed using the LBD model for electron-electron collisions, which is found to substantially increase particle losses. This result is consistent with the recent work in Ref.~[\onlinecite{rosen:2025}], where a method to improve the LBD model for more accurate loss predictions is proposed.

The simulations illustrated in Fig.~\ref{Fig_BASM_schematic} use plasma and magnetic mirror parameters inspired by recent numerical studies of the WHAM experiment~\cite{endrizzi:2023, francisquez:2023} (see Table~\ref{tab:WHAM}). The electron species is initialized with a Maxwellian distribution characterized by $n_0=\SI{e19}{\m^{-3}}$ and $T_e = \SI{940}{\eV}$. The magnetic mirror ratio is $R_m = 32$, and the length of the magnetic trap is $L_\parallel = \SI{2}{\m}$, corresponding to a weakly collisional regime with $\nu_e V_{Te} / L_\parallel = 2.6\times10^{-3}$. Here, $V_{Te} = \sqrt{2T_e/m_e}$ is the electron thermal velocity, and $m_e$ is the electron mass. Note that for the 0D-2V BASM model, the trap length parameter enters the simulation only through the sink term in Eq.~(\ref{SinkTerm}), where it controls the rate of prompt particle losses from the loss-cone region. No sources are included in the simulations, $\Gamma_{src} = 0$, and the electron distribution function evolves according to Eq.~(\ref{BASMeq}). The velocity grid resolution is given by $N_{\vpll} = 1024$, $N_\mu = 1024$, and the velocity domain extent corresponds to $-5 {V}_{Te} \le \vpll \le 5 {V}_{Te}$ and $0 \le \mu \le 12 T_0 / B_0$, where $B_0$ is the magnetic field at the trap center. The time step is set to $\Delta t = \SI{3.3}{\micro \second}$, which corresponds to a collisional CFL number of $1.5\times10^3$ for the Fokker-Planck operator. Numerical convergence of the simulation results with respect to the grid resolution, domain extent, and time step is confirmed. 

\begin{table}
\caption{\label{tab:WHAM} Reference parameters used in COGENT simulations.  }
\begin{ruledtabular}
\begin{tabular}{ccc}
Parameter&Description&Value\\
\hline
\rule{0pt}{2.5ex}
$n_0$ & Initial plasma density & $\SI{e19}{\m^{-3}}$\\
$T_0$ & Initial ion temperature & $\SI{8.361}{\keV}$\\
$T_e$ & Electron temperature & $\SI{940}{\eV}$\\
$m_i$ & Ion mass & $2m_p$\\
$R_m$ & Mirror ratio & 32\\
$L_\parallel$ & Trap length & \SI{2}{\m}\\
$B_0$ & Magnetic field at the mirror center & \SI{0.5}{\tesla}\\
$E_b$ & Ion beam energy & \SI{25}{\keV}\\
$\theta_{inj}$ & Beam injection angle & \SI{45}{\degree}\\
$\Gamma_b$ & Ion beam intensity & \SI{2.3e15}{\s^{-1}\cm^{-3}}\\
\end{tabular}
\end{ruledtabular}
\end{table}

Figure~\ref{Fig_BASM_schematic}(c) shows the time history of the electron confinement time. Simulation results obtained using the full nonlinear FP operator for e-e collisions are compared with those from counterpart simulations, where e-e collisions are modeled as scattering off a fixed Maxwellian background corresponding to the initial electron distribution. Following initial agreement over a shorter (collisional) timescale, the full FP model yields an increasing confinement time compared to the fixed-background model over a longer (transport) time scale. This behavior can be explained as follows: particles that are scattered in the loss-cone region and lost from the trap have significantly higher-than-average energy for $ e \Phi_m/T_e \gg 1$. As the energy-conserving collisions modeled by the full FP operator thermalize the electron distribution, its temperature decreases. This leads to improved electrostatic confinement, quantified by $e\Phi_m/T_e$, and thus to an increase in the confinement time. We also note small changes in the characteristic electron collision frequency in Eq.~(\ref{nuCohen}) due to the decreasing electron density and temperature. However, these effects are subdominant compared to the exponential sensitivity of the confinement time to the electrostatic potential. In contrast, when e-e collisions are modeled as collisions with a fixed Maxwellian background, energy is not conserved, and the electron temperature remains close to that of the background distribution, i.e., the initial Maxwellian. Additionally, the density in Eq.~(\ref{nuCohen}) corresponds to the background Maxwellian and therefore remains fixed. As a result, a steady-state solution for $\tau_{c,e}$ is observed. It is instructive to note that the approximate analytical solution in Eq.~(\ref{LossesAnalytic}) assumes collisions with a fixed Maxwellian background. In this context, Figure~\ref{Fig_BASM_schematic}(b) presents numerical simulation results obtained using the fixed-background model for both the FP and LBD e-e collision operators. In the latter case, the fixed-background model assumes constant density and temperature in Eqs. (\ref{LBDOp}) and (\ref{nuLBD}) corresponding to the initial Maxwellian.    

\subsection{High-energy beam relaxation}

\begin{figure*}
\includegraphics[width=0.8\textwidth]{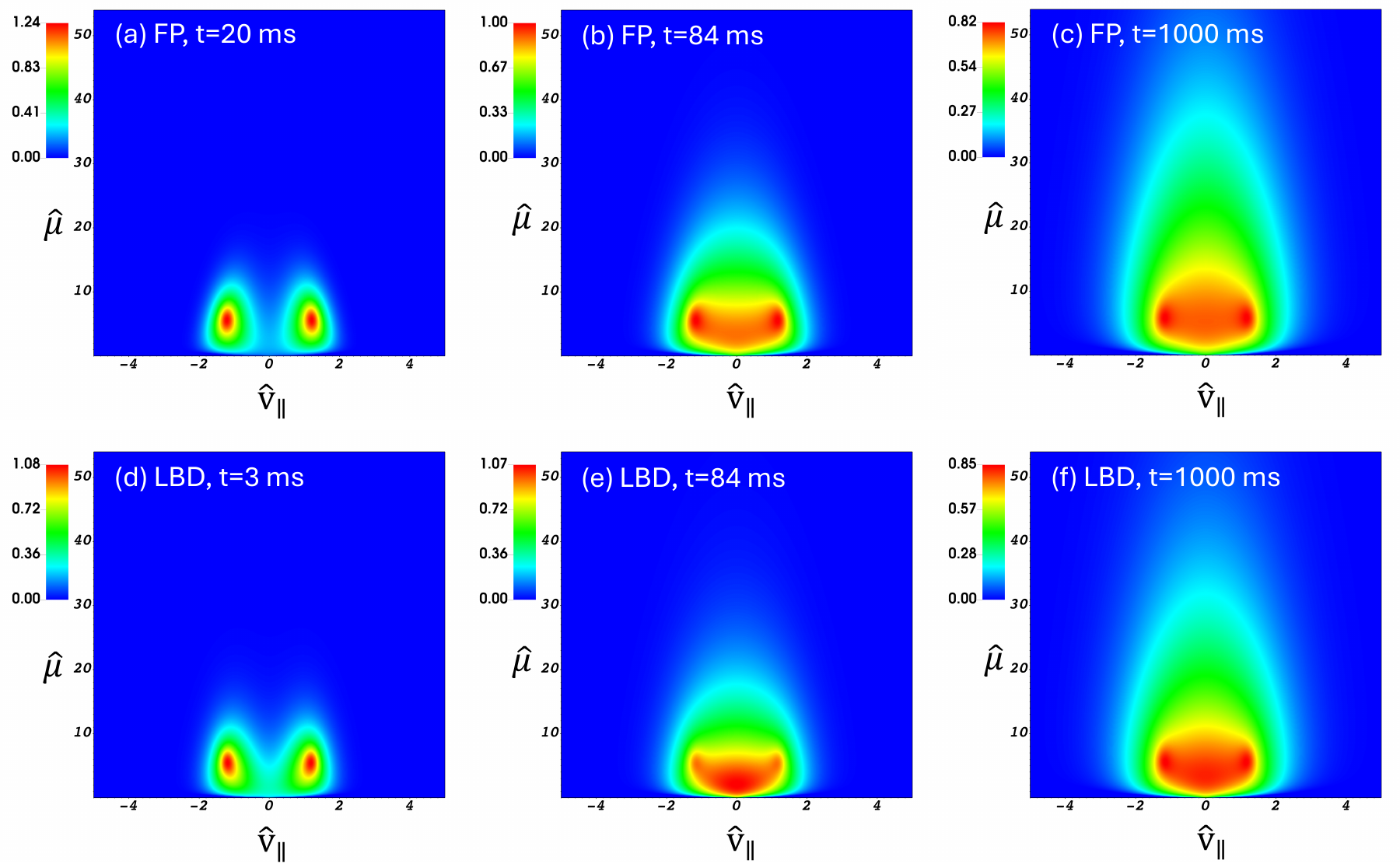}
 \caption{High-energy beam relaxation obtained from 0D–2V BASM model simulations. The numerical results using the Fokker–Planck collision model [frames (a)–(c)] are compared with corresponding simulations using the LBD collision model [frames (d)–(f)]. Plotted is the normalized ion distribution function, ${\hat f}_i = (2T_0/m_i)^{3/2} \pi f_i /n_0$.}
 \label{Fig_2V_FP_LBD}
\end{figure*}

The BASM model is used here to provide insight into numerical simulations of high-energy beam fueling of a magnetic mirror system. In particular, we use it to examine the numerical convergence of the velocity-space discretization algorithms, which are challenged by the presence of steep gradients at the loss-cone boundary for a weakly collisional mirror. For simplicity, only a deuterium background distribution and beam ions are included in the simulation, and we omit the effects of ion-electron (cooling) collisions and self-consistent electric fields. As an illustrative example, we consider physical parameters inspired by recent numerical studies of the WHAM experiment~\cite{endrizzi:2023, francisquez:2023}: $B_0 = \SI{0.5}{\tesla}$, $R_m = 32$, $L_\parallel = \SI{2}{\meter}$, $n_0 = \SI{e19}{\meter^{-3}}$, $T_0=\SI{8.361}{\keV}$, $m_i = 2m_p$, $E_b = \SI{25}{\keV}$, $\theta_\mathrm{inj} = \SI{45}{\degree}$. Here, $n_0$ and $T_0$ correspond to the ion density and temperature of the initial Maxwellian background distribution, $E_b$ is the energy of beam ions, and $\theta_\mathrm{inj}$ is the beam injection angle. Beam fueling is described by the source term on the right-hand side of Eq.~(\ref{BASMeq}), where we arbitrarily adopt the following shape function
\begin{align}
 \Gamma_{src}  &= \Gamma_0 \exp \left( - \frac{ (\vpll - V_{b,\parallel})^2 +  \left( \sqrt{ \frac {2\mu B_0}{m_i}} - V_{b,\perp} \right)^2 }{2 T_b / m_b} \right) \nonumber \\
    & + \Gamma_0 \exp \left( - \frac{ (\vpll + V_{b,\parallel})^2 +  \left( \sqrt{ \frac {2\mu B_0}{m_i}} - V_{b,\perp} \right)^2 }{2 T_b / m_b} \right).
  \label{BeamSrc}
\end{align}
Here, $V_{b,\parallel} = V_{b,\perp} = \sqrt{E_b/m_i}$, $T_b = \SI{200}{\eV}$, and the normalization constant $\Gamma_0$ is determined such that the beam intensity, $\Gamma_b =  ( 2 \pi / m_i) \int \Gamma_{src} B d\vpll d{\mu} =  \SI{2.3e15}{\s^{-1} \cm^{-3}} $, corresponds to a plasma density of approximately $\SI{3e19}{\m^{-3}}$ after \SI{10}{\ms} consistent with the analysis in Ref.~[\onlinecite{endrizzi:2023}].

\begin{figure}
\includegraphics[width=\linewidth]{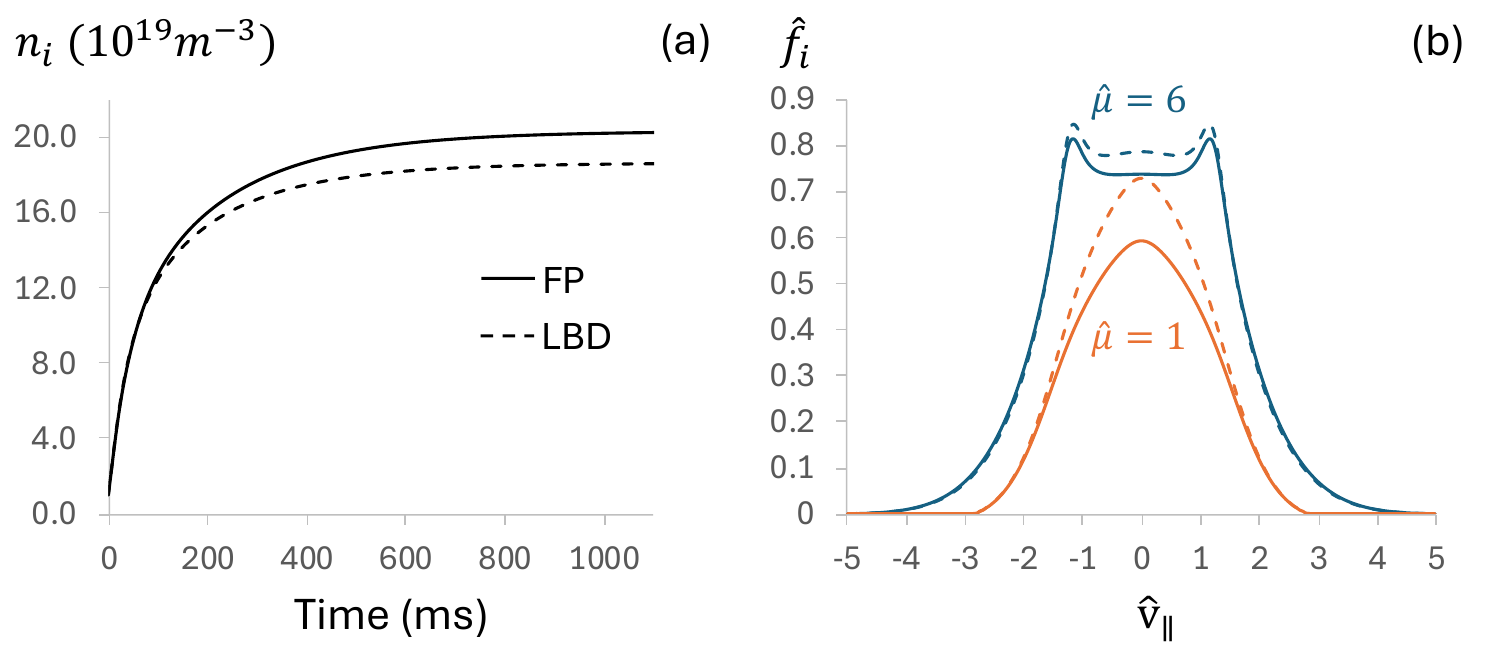}
\caption{Comparison of the Fokker-Planck (solid curves) and LBD (dashed curves) collisional models in the 0D-2V BASM-model simulations. Shown are (a) time history of the ion density; and (b) the $\vpll$-lineouts extracted from the normalized ion distribution function, ${\hat f}_i = (2T_0/m_i)^{3/2} \pi f_i /n_0$, for $\widehat \mu = 1$ and $\widehat \mu = 6$, at 1000 ms. }
\label{Fig_2V_timeHistory}
\end{figure}

Figures~\ref{Fig_2V_FP_LBD} and \ref{Fig_2V_timeHistory} show the results of numerical simulations obtained using the full nonlinear Fokker-Planck (FP) operator and the model LBD operator to describe ion-ion collisions. A nearly steady-state particle density is reached after approximately \SI{1000}{\ms} [see Fig.~3(a)], as beam fueling is balanced by collisional losses, i.e., scattering into the loss-cone region. A pronounced difference in the phase-space solution is evident between the collisional models, especially during the evolution stage [c.f., Fig. \ref{Fig_2V_FP_LBD}(b) and \ref{Fig_2V_FP_LBD}(e)],   emphasizing the importance of using a detailed collision operator.

\begin{figure}
\includegraphics[width=\linewidth]{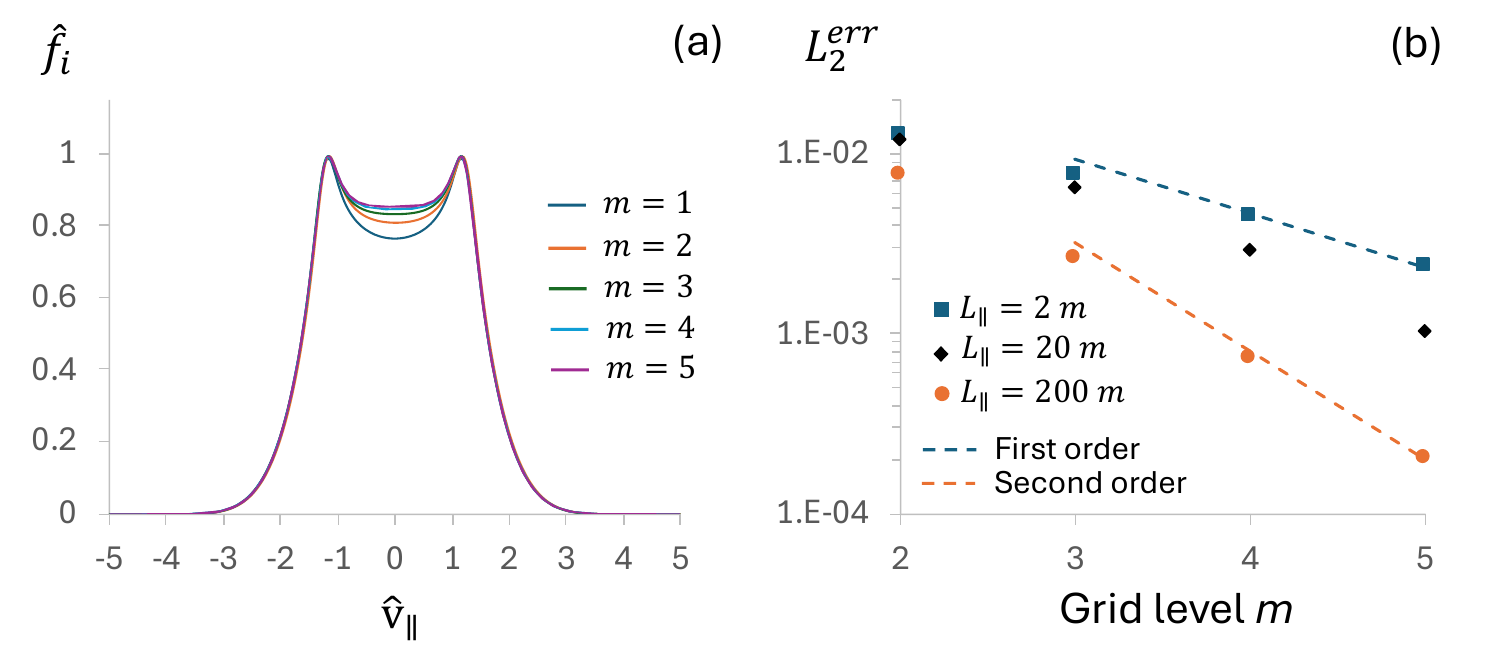}
\caption{Numerical convergence of the 0D-2V BASM model solution at \SI{84}{\ms}. Shown are (a) the $\vpll$-lineouts extracted from the normalized ion distribution function, ${\hat f}_i = (2T_0/m_i)^{3/2} \pi f_i /n_0$, for $\widehat \mu = 6$; and (b) Richardson extrapolation analysis of the numerical convergence for different values of the trap length, $L_\parallel$. Grid levels $m=(2,3,4,5)$ correspond to the successful refinements of the coarse grid $(N_{\vpll},N_\mu)=(128,192)$ by factors of 2, 4, 8, and 16 in both velocity dimensions, respectively. }
\label{Fig_2V_converg}
\end{figure}

For the simulations illustrated in Figs.~\ref{Fig_2V_FP_LBD} and \ref{Fig_2V_timeHistory}, the velocity grid resolution is given by $(N_{\vpll}, N_\mu) = (512,768)$, and the velocity domain extent corresponds to $-5 \le \vpllhat \le 5$ and $0 \le \widehat{\mu} \le 54$, where $\vpllhat = \vpll / V_{T0}$, $\widehat{\mu} = 4 B_0 \mu /T_0$, and $V_{T0}=\sqrt{2T_0/m_i}$. Numerical convergence studies with respect to velocity grid resolution are shown in Fig.~\ref{Fig_2V_converg} for the case of the Fokker-Planck collisional model. Figure~\ref{Fig_2V_converg}(a) illustrates a lineout of the normalized distribution function, ${\hat f}_i = (2T_0/m_i)^{3/2} \pi f_i /n_0$, along the $\vpll$-coordinate, obtained from simulations using a coarse velocity grid $N_{\vpll} = 128$, $N_\mu = 192$ and its successful refinements by factors of 2, 4, 8, and 16 in both velocity dimensions. Although convergence of the numerical solution is apparent, comparison of values at $\vpll = 0$ suggests that the convergence is only first-order, despite the use of second-order discretization in the numerical implementation of Eq.~(\ref{BASMeq}). First-order convergence is supported by a more rigorous analysis using the Richardson extrapolation method. Figure~\ref{Fig_2V_converg}(b) plots the $L^2$ norm of the difference between the numerical solution ${\hat f}_i$ obtained at a given grid spacing, $h$, and that at half the spacing, $h/2$, i.e., $L_2^{err}= \| {\hat f}_{h} - {\hat f}_{h/2} \| $.  

The degraded convergence can be explained as follows. For a weakly collisional regime, $\nu_{ii} L_\parallel / V_{T} \ll 1$, the solution to Eq.~(\ref{BASMeq}) is distinguished by the presence of a narrow-layer region at the loss-cone boundary that mediates the transition between the confined and loss-cone regions. Here, $V_{T}$ is the characteristic thermal velocity. The characteristic width of this region, $\delta \mathrm{v}$ is determined from the balance between the sink term (prompt losses) and the collisional diffusion, $C_{ii} [f] \sim \Gamma_{sink}[f]$, and is given by     
\begin{equation}
\delta \mathrm{v}  \sim \left( V_{T} \nu_{ii} L_\parallel \right) ^ {1/2}.
  \label{DeltaV}
\end{equation}
If the narrow layer in Eq.~(\ref{DeltaV}) is not sufficiently resolved by the computational grid, degradation in numerical convergence properties can be expected. In the asymptotic limit, a weakly collisional solution can be obtained by imposing a zero Dirichlet boundary condition directly at the loss-cone boundary. However, not resolving the transition length effectively corresponds to imposing a zero boundary condition at the nearest cell center below the loss-cone boundary, rather than at its actual location. A cell-size error in the boundary condition specification can degrade convergence to first-order accuracy. It is instructive to note that embedded boundary methods~\cite{johansen:1998,thacher:2023} can provide high-order accurate solutions for weakly collisional formulations that impose a zero Dirichlet boundary condition at a loss-cone boundary intersecting the grid cells.

For the numerical solution in Fig.~\ref{Fig_2V_FP_LBD}(b), which is used for the convergence studies in Fig.~\ref{Fig_2V_converg}, we have $T=\SI{21}{\keV}$, $n_i = \SI{12e19}{\m^{-3}}$, and $\delta \mathrm{v} /V_T \sim 6.3\times10^{-3}$, where Eq.~(\ref{nuLBD}) is used to estimate the ion-ion collision frequency in Eq.~(\ref{DeltaV}). At the same time, our most refined grid $(N_{\vpll},N_\mu)=(2048,3072)$ with $\Delta \vpll /V_T = 3.0\times 10^{-3}$ and $\Delta \vperp /V_T \approx 5.0\times 10^{-3}$ does not sufficiently resolve the transition layer, therefore degraded convergence may be expected. Here, $V_T=\sqrt{2T/m_i}$, and we used $\Delta \mu = m \vperp \Delta \mathrm{v} / B_0 \approx m V_T \Delta \mathrm{v} / \sqrt{R_m} B_0 $ at the loss-cone boundary. To further investigate this issue, we consider additional simulations in which the sink term is arbitrarily reduced by factors of 10 and 100 [see Fig.~\ref{Fig_2V_converg}(b)]. For instance, this can be achieved by increasing the trap length, which leads to a higher collisionality parameter, $\nu_{ii} L_\parallel / V_{T}$, and a wider transition layer, $\delta \mathrm{v}$. Second-order numerical convergence is readily observed in the case with the highest collisionality. For the medium-collisionality case, the convergence improves from first- to second-order as the grid resolution increases and the transition layer becomes resolved.   

\section{Fully kinetic (1D-2V) simulations}

In this section, we investigate the performance of our implicit numerical algorithms for the modeling of parallel plasma transport in a weakly collisional magnetic mirror. As an illustrative example, we consider the problem of high-energy beam fueling of a WHAM mirror system [see Sec.~II~C] that now includes a magnetic field profile, ion parallel streaming, and a self-consistent parallel electric field determined from the Boltzmann electron model. 

\subsection{Simulation model}

Assuming the paraxial approximation with $B_z \gg B_r$ and $\nabla_\parallel \approx \partial / \partial z $, the time evolution of the ion distribution function, $f_i (z, \vpll, \mu)$, within a magnetic flux tube is governed by
\begin{equation}
\frac{\partial f_i}{\partial t} + \vpll \frac{\partial f_i}{\partial z} + \frac{\partial }{\partial \vpll} \left( \left[ \frac{q_i}{m_i} E_z - \frac{\mu}{m_i} \frac{\partial B_z}{\partial z} \right] f_i \right) = C_{ii} [f_i] + \Gamma_{src},
\label{VlasovEq}
\end{equation}
where $E_z$  is the parallel electric field, and $q_i$ and $m_i$ denote the ion species charge and mass, respectively. In deriving Eq.~(\ref{VlasovEq}), we used the fact that the Jacobian of the transformation to $(\vpll, \mu)$ velocity coordinates, $J_{\bf v} = (2\pi/m_i) B$, cancels the spatially varying part of the Jacobian associated with a magnetic-flux-surface coordinate, $J_{\bf x} \propto 1/B$. As a result, their product reduces to a constant factor, $J_{\bf v} J_{\bf x} \propto 2\pi/m_i$. The full nonlinear Fokker-Planck collision model is used for like-species ion-ion collisions, $C_{ii}[f]$. Collisions with electrons are not included in the present work, for simplicity. The electric field is obtained from the quasi-neutrality condition coupled to the Boltzmann electron model:
\begin{equation}
E_z = - \frac{T_e}{en_i} \frac{\partial n_i}{\partial z},
\label{adiabEl}
\end{equation}
where  $n_i = (2\pi/m_i) \int d\vpll d\mu f_i B$ is the ion density, and $T_e$ is a uniform electron temperature. 

\begin{figure}
\includegraphics[width=0.4\textwidth]{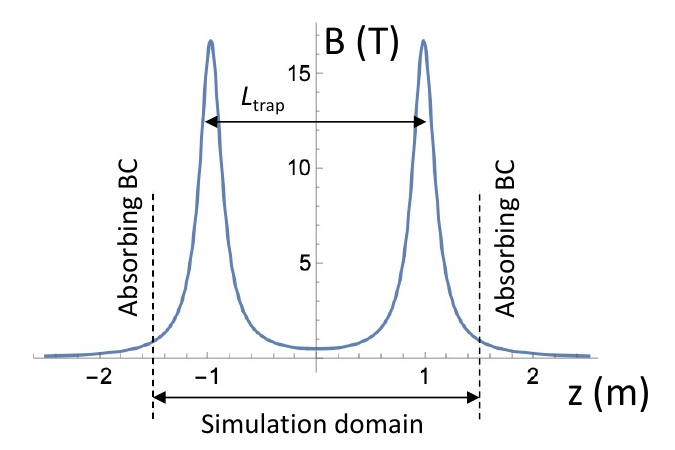}
 \caption{Magnetic geometry used in 1D-2V fully kinetic simulations. }
 \label{Fig_1D2V_schematic}
\end{figure}

An analytical approximation to the WHAM magnetic geometry (see Fig.~\ref{Fig_1D2V_schematic}) follows the approach in Ref.~[\onlinecite{francisquez:2023}]: 
\begin{equation}
B_z(z) = \frac{\overline B}{\pi \gamma} \left\{ \left[ 1 + \left( \frac{Z-Z_m}{\gamma} \right)^2 \right]^{-1} + \left[ 1 + \left( \frac{Z+Z_m}{\gamma} \right)^2 \right]^{-1} \right \},
\label{Bwham}
\end{equation}
where $\overline B = \SI{6.5}{\tesla}$, $\gamma=0.124$, $Z_m=\SI{0.98}{\m}$. The simulation domain extent is given by $\SI{-1.5}{\m}<z<\SI{1.5}{\m}$, and a particle-absorbing boundary condition is imposed in the z-direction. The source function in Eq.~(\ref{VlasovEq}) follows our choice in Eq.~(\ref{BeamSrc}), where a spatially uniform normalization constant, $\Gamma_0$,  is now replaced with a spatially varying function $\Gamma_0 \exp(-z^2/L_b^2)$, where $L_b = \SI{0.2}{\m}$. The normalization constant $\Gamma_0$ is determined from the condition that it matches the particle production rate used in the BASM simulations in Sec.~II.~C, i.e.,  $\Gamma_b =  \int J_{\bf v} J_{\bf x} \Gamma_{src} d\vpll d{\mu} dz / \int J_{\bf x} dz= \SI{2.3e15}{\s^{-1} \cm^{-3}}$. Here, the integral along the $z$-coordinate is evaluated between the mirror throat points. The deuterium ion species, $m_i = 2 m_p$, is initialized with a Maxwellian distribution, which has a uniform temperature $T_0 = \SI{8.361}{\keV}$ and a nearly flat-top density profile with a steep decrease near the magnetic plugs, $n_i(t=0) = {\overline C} n_0 [\tanh\left( (z_0 - z)/ {\overline L}\right) + \tanh\left( (z_0 + z)/ {\overline L}\right)]$, where $n_0 = \SI{e19}{\m^{-3}}$, $z_0 = \SI{0.75}{\m}$, $\overline L = \SI{0.1}{m}$, and the normalization coefficient $\overline C = 0.52$ is chosen to match the initial number of particles in the BASM simulations in Sec.~II~C. Following the simulations in Ref.~[\onlinecite{francisquez:2023}], we set $T_e = \SI{940}{eV}$.

Due to the substantial computational cost of the numerical simulations, we run them only for 20 ms, which roughly corresponds to the duration of the WHAM beam source. The results from the 0D-2V BASM model show that the phase-space solution is much more compact in velocity space at $t = \SI{20}{\milli\second}$ compared to the final steady-state solution [cf. Figs.~\ref{Fig_2V_FP_LBD}(a) and \ref{Fig_2V_FP_LBD}(c)]. Therefore, a smaller velocity-domain extent is used for the 1D-2V simulations: $-2.5 \le \vpllhat \le 2.5$ and $0 \le \widehat{\mu} \le 18$, where $\vpllhat = \vpll / V_{T0}$, $\widehat{\mu} = 4 B_0 \mu /T_0$, $B_0 = \SI{0.5}{\tesla}$, and $V_{T0}=\sqrt{2T_0/m_i}$. The baseline grid resolution is given by $(N_z, N_{\vpll}, N_\mu) = (256, 128, 192)$.

To illustrate the strong disparity in time scales, we consider the initial background plasma parameters as a reference, for which the characteristic ion transit time is given by $\tau_\parallel = L_{\mathrm{trap}}/V_{T0} = \SI{2.2}{\micro\s}$, while the characteristic ion-ion collision time is $\tau_{ii} = 1/\nu_{ii} = \SI{100}{\ms}$. Here, $L_{\mathrm{trap}}$ is the distance between the mirror throat points, and the collision frequency $\nu_{ii}$ is estimated using Eq.~(\ref{nuLBD}). A time step for stable explicit simulations is limited by the CFL constraint, $\Delta t_{\text{CFL}} \lesssim \min {\left( \Delta t_\parallel ^{\text{CFL}} , \Delta t_{\nabla B} ^{\text{CFL}} \right)}$, where $\Delta t_\parallel \sim \Delta z / \vpll^{max} = \SI{5.2}{\ns}$,  $\Delta t_{\nabla B}^{\text{CFL}} \sim {m_i \Delta \vpll}/({\mu_{max} | \nabla B |_{max}})=\SI{0.1}{\ns}$. It is clear that explicit time integration of a weakly collisional mirror plasma on the ion-ion collisional timescale becomes prohibitively expensive, and implicit time integration is required for first-principles kinetic modeling. 

\begin{figure*}
\includegraphics[width=0.8\textwidth]{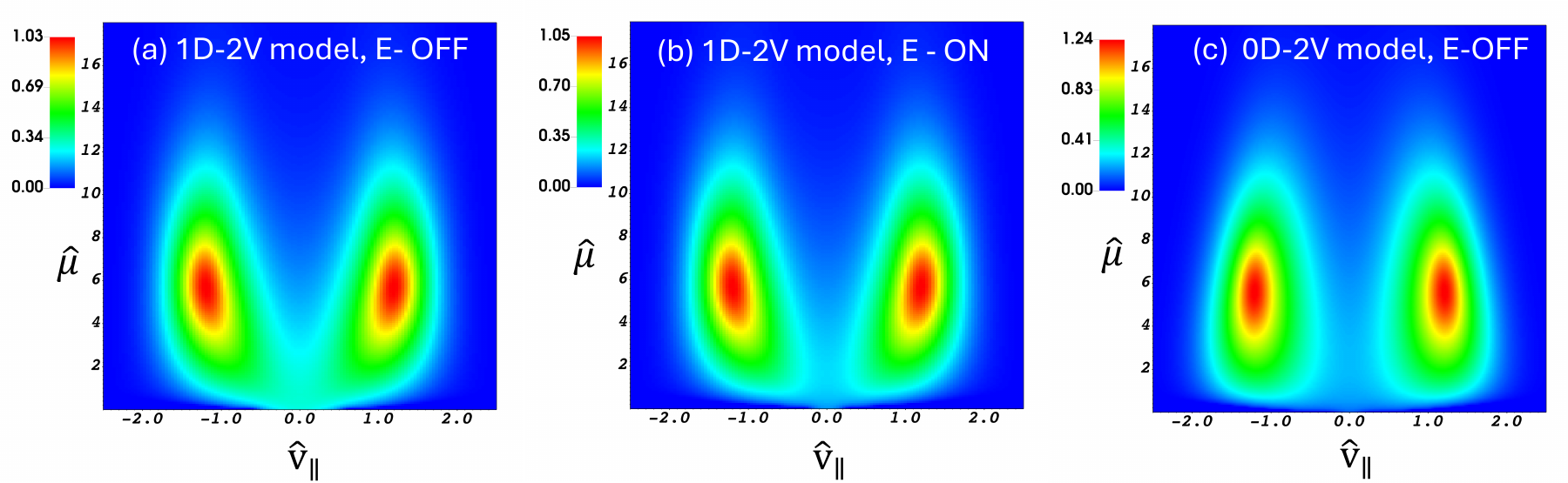}
 \caption{Normalized ion distribution function, ${\hat f}_i = (2T_0/m_i)^{3/2} \pi f_i /n_0 $, evaluated at the trap center at $\SI{20}{\ms}$. The results are shown for (a) fully kinetic 1D-2V simulations in the absence of E-fields; (b) fully kinetic 1D-2V simulations including a self-consistent E-field specified by the Boltzmann electron model; and (c) 0D-2V BASM-model simulations in the absence of E-fields. Frame (c) corresponds to a zoom-in of Fig.~\ref{Fig_2V_FP_LBD}(a).}
 \label{Fig_1D2V_vparmu}
\end{figure*}

\begin{figure}[b]
\includegraphics[width=\linewidth]{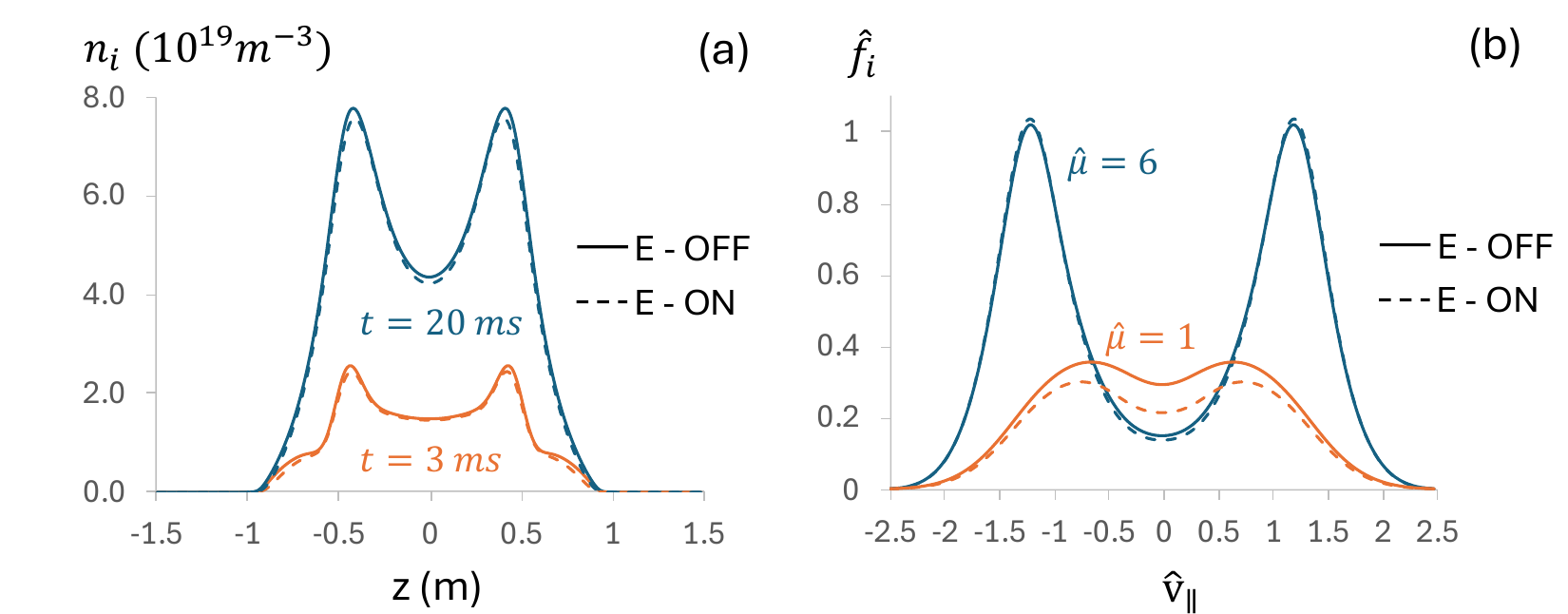}
 \caption{Results of the fully kinetic 1D-2V simulations with (solid curves) and without (dashed curves) E-fields. Shown are (a) the ion density profile evaluated at \SI{3}{\ms} (orange) and  \SI{20}{\ms} (blue); and (b) the $\vpll$-lineouts extracted from the normalized ion distribution function, ${\hat f}_i$, at the center of the trap, shown for $\widehat \mu = 1$ and $\widehat \mu = 6$, at 20 ms.}
 \label{Fig_1D2V_lineouts}
\end{figure}

Here, the feasibility of multiscale collisional mirror transport simulations is explored using a second-order implicit–explicit (IMEX) time integration scheme, which enables implicit treatment of selected stiff terms. Specifically, the nonlinear Fokker-Planck operator, associated with the long ion-ion collisional time scale, is treated explicitly, whereas the Vlasov operator, involving fast advection in both the $z$ and $\vpll$ directions, is included as an implicit term. To handle nonlinearities in the Vlasov operator due to the presence of a self-consistent electric field, the Jacobian-free Newton-Krylov (JFNK) approach is employed~\cite{knoll:2004} with the GMRES method used as a Krylov solver~\cite{saad:2003}. Although this approach alleviates the need to form or store the Jacobian matrix corresponding to the linearized system, it requires preconditioning for efficiency. In this work, we construct the preconditioner operator by neglecting the $\partial E_z / \partial f_i$ semi-dense contribution to the actual Jacobian operator. Physically, this means that our preconditioner operator does not capture collective modes involving the self-consistent electric field response, such as the plasma sound wave, $\omega =[(\gamma T_i + T_e)/m_i]^{1/2} k_\parallel$. Here, $k_\parallel$ is a parallel wave vector and $\gamma$ is an ion adiabatic index. We, however, note that for mirror systems, e.g., WHAM, where the ion temperature is much greater than the electron temperature, $T_i \gg T_e$, corrections to the linear-response Jacobian due to the self-consistent electric field model in Eq.~(\ref{adiabEl}) may be subdominant to the "fixed-$E_z$" ion advection contribution.     

Neglecting couplings between ion distribution function and the electric field in the Jacobian preconditioner, yields a 3D (1D-2V) non-symmetric and indefinite sparse matrix for the discretized "fixed-$E_z$" advection operator. COGENT solves this linear system using the Approximate Ideal Restriction (AIR) option in the BoomerAMG algebraic multigrid solver contained in the Hypre linear solver library~\cite{Hypre}. The use of multigrid methods to solve nonsymmetric indefinite systems has historically been highly problematic, but the recent development of the AIR approach~\cite{manteuffel:2018, manteuffel:2019}, including several variants, provides a way to extend the benefits of multigrid algorithms beyond the symmetric, positive-definite systems for which they are more commonly used. To further enhance the efficiency of the implicit Vlasov solver, COGENT allows the preconditioner to be defined using a lower-order discretization than that used for the Vlasov operator itself. In the weakly collisional regime, $\tau_\parallel \ll \tau_{ii}$, the collisional term in Eq.~(\ref{VlasovEq}) is subdominant to the Vlasov term. Therefore, a numerical scheme with high spatial accuracy must be employed for the discretization of the Vlasov operator in order to mitigate truncation errors and minimize numerical pollution. In this work, we use a fifth-order upwind discretization (UW5) for the Vlasov operator in Eq.~(\ref{VlasovEq}), whereas the preconditioner operator is constructed using a first-order (UW1) advection scheme. This lower-order preconditioner yields a sparser matrix, for which the most robust AMG solver performance is observed, while also maintaining efficiency even for relatively large time steps. 

\subsection{Zero E-field simulations }

\begin{figure*}
\includegraphics[width=0.8\textwidth]{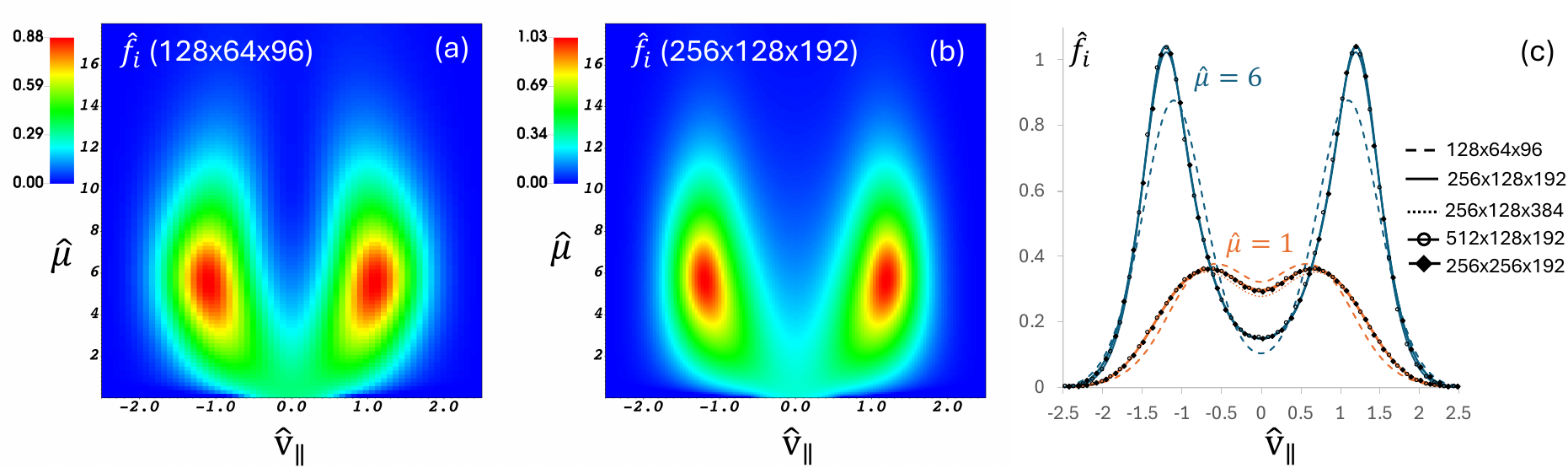}
 \caption{Numerical convergence of the fully kinetic 1D-2V simulations without E-fields employing an UW5 discretization scheme for the Vlasov operator. The normalized ion distribution function is evaluated at the trap center at $\SI{20}{\ms}$.}
 \label{Fig_1D2V_convergence}
\end{figure*}

\begin{figure*}
\includegraphics[width=0.8\textwidth]{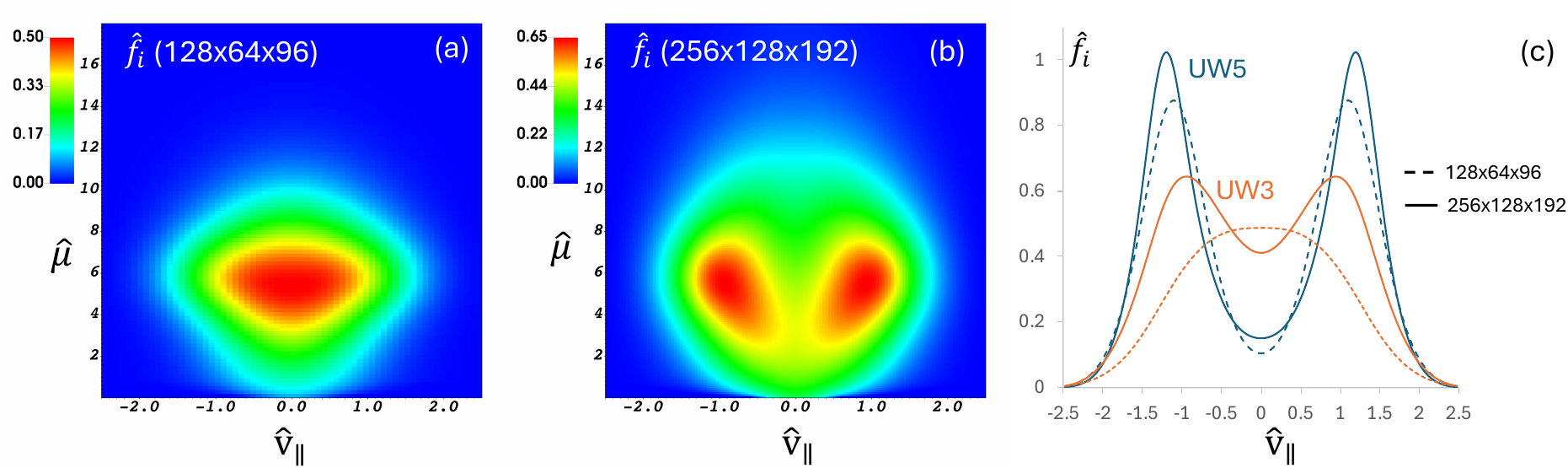}
 \caption{Numerical convergence of the fully kinetic 1D-2V simulations without E-fields employing an UW3 discretization scheme for the Vlasov operator. The normalized ion distribution function is evaluated at the trap center at $\SI{20}{\ms}$. Frame (c) compares the $\vpll$-lineouts at $\widehat \mu = 6$ for the UW3 (orange) and UW5 (blue) discretization schemes.}
 \label{Fig_1D2V_UW3}
\end{figure*}

Here, simplified 1D-2V simulations are performed in the absence of electric-field effects. In this case, the preconditioner operator is consistent with the physical operator (modulo the difference in discretization order) allowing us to investigate the maximal benefits of our implicit algorithms. This simplified setup is also used to evaluate the convergence properties and numerical pollution associated with our spatial discretization.

The results of numerical simulations using the baseline grid resolution $(N_z, N_{\vpll}, N_\mu)=(256,128,192)$ are shown in Fig.~\ref{Fig_1D2V_vparmu}(a) for the normalized distribution function ${\hat f}_i = (2T_0/m_i)^{3/2} \pi f_i /n_0$ evaluated at the trap center. Qualitative agreement with the results of the BASM model is observed [cf. Fig.~\ref{Fig_1D2V_vparmu}(a) and Fig.~\ref{Fig_1D2V_vparmu}(c)]; however the first-principles kinetic model captures additional features, e.g., the magnetic field profile. In particular, the peaks in the ion density profiles in Fig.~\ref{Fig_1D2V_lineouts}(a) correspond to the beam turning points.

The numerical convergence studies of the final-state solution at \SI{20}{\ms} are illustrated in Figure~\ref{Fig_1D2V_convergence}. Performing a mathematically rigorous analysis involving multiple levels of grid refinement in all phase-space dimensions, as done for the 0D-2V BASM model in Sec.~II~C, is too computationally expensive. Therefore, here we consider the following set of computational grids: a coarse grid (128,64,96), a baseline grid (256,128,192), and three semi-fine grids with double the number of grid points in each individual direction, i.e., (512,128,192), (256, 256,192), and (256,128,384). To gain insights into numerical convergence, we examine two $\vpll$-lineouts in the ($\vpll,\mu$) velocity space, taken at the center of the trap and corresponding to $\widehat \mu = 6$ and $\widehat \mu = 1$, which represent the high-energy beam and thermal background regions, respectively [see Fig.~\ref{Fig_1D2V_convergence}(c)]. Notably, in the beam region characterized by narrow peaks, numerical errors are more pronounced between the baseline and coarse grids compared to the background region, which exhibits a more diffuse profile. The results from simulations with the semi-fine grids show only a small difference relative to the baseline resolution in both regions. 

As mentioned earlier, it is important to use a high-order (UW5) spatial discretization of the Vlasov operator to minimize numerical pollution. To further illustrate this point, we compare the results of our baseline and coarse grid simulations using the UW5 scheme with those obtained using a lower-order UW3 discretization (see Fig.~\ref{Fig_1D2V_UW3}). It is readily apparent that the excessive numerical diffusion produced by the UW3 scheme over long timescales, $t \gg \tau_\parallel$, leads to a completely spurious solution for the coarse grid case [see Fig.~\ref{Fig_1D2V_UW3}(a)]. While distinct beam features begin to emerge with the more resolved baseline grid [see Fig.~\ref{Fig_1D2V_UW3}(b)], the solution remains far from being quantitatively converged. In contrast, the UW5 scheme yields a converged velocity-space solution at the same (baseline) resolution.    

\begin{table}
\caption{\label{tab:IMEX} Performance of COGENT’s IMEX scheme for simulations with the baseline grid resolution $(N_z, N_{\vpll}, N_\mu) = (256, 128, 192)$ using 512 cores.  }
\begin{ruledtabular}
\begin{tabular}{ccc}
Parameter&$E$ -- OFF &$E$ -- ON\\
\hline
\rule{0pt}{2.5ex}
Implicit time step / wall-time per step & $\SI{4.7}{\micro\s}/ \SI{7.5}{\s}$ & $\SI{4.7}{\micro\s}/ \SI{9.5}{\s}$\\
Explicit time step / wall-time per step & $\SI{0.19}{\nano\s} / \SI{0.9}{\s}$ & $\SI{0.19}{\nano\s}/\SI{1.0}{\s}$\\
Number of JFNK iterations (1.0e-4 tol.) & $\approx 55$& $\approx 70$\\
\end{tabular}
\end{ruledtabular}
\end{table}

We now discuss the computational efficiency of our IMEX time integration approach for the 1D-2V kinetic simulations (see Table~\ref{tab:IMEX}). The stable time step is limited by the CFL constraint associated with the explicitly treated Fokker-Planck operator. For the baseline grid resolution, the time step is initially limited by $\Delta t_\mathrm{init}^{CFL} \approx \SI{20}{\micro\second}$ and decreases to $\Delta t_\mathrm{fin}^{CFL} \approx \SI{4.7}{\micro\second}$ by the end of the simulation at $t_\mathrm{fin} = \SI{20}{\ms}$, as the collisional time scales shorten due to increasing ion density. The baseline grid simulation uses 512 cores of the NERSC system~\cite{NERSC} and requires approximately $\SI{20}{\s}$ and $\SI{7.5}{\s}$ per $\Delta t_\mathrm{init}^{CFL}$ and $\Delta t_\mathrm{fin}^{CFL}$ steps, respectively. The JFNK solver tolerance is set to $10^{-4}$, and it takes about 100 and 55 JFNK iterations for $\Delta t_\mathrm{init}^{CFL}$ and $\Delta t_\mathrm{fin}^{CFL}$, respectively. For comparison, the stable time step for a fully explicit time integration scheme is limited by $\Delta t_{\nabla B}^{\text{CFL}} \approx \SI{0.19}{\nano\second}$, and each step requires approximately $\SI{0.9}{\s}$. Consequently, the IMEX simulations achieve a speedup of about $5 \times 10^3$ in the initial stage and $3 \times 10^3$ in the final stage. It is, however, instructive to note that the computational cost of an explicit simulation is dominated by the evaluation of the Fokker--Planck operator; disabling it reduces the wall time to about $\SI{0.15}{\s}$ per step. Thus, explicit simulations can be accelerated by about a factor of 6 if the Rosenbluth potentials are evaluated infrequently.

\begin{figure*}
\includegraphics[width=0.9\textwidth]{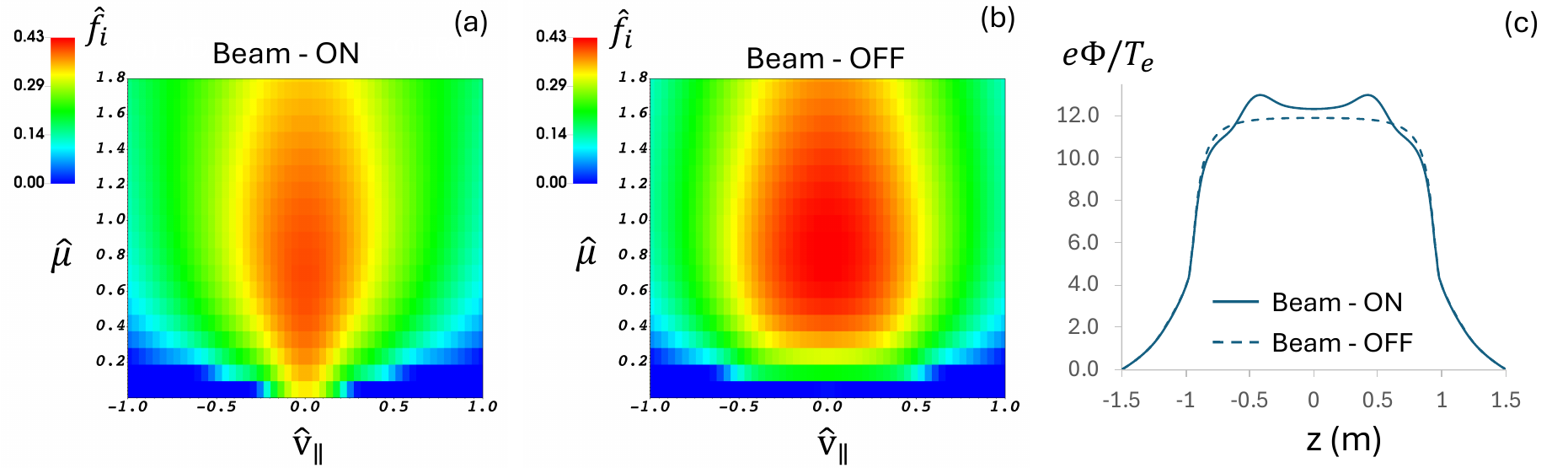}
 \caption{Role of the sloshing-ion potential well in low-energy ion confinement. Shown are results from 1D-2V fully kinetic simulations with (a) and without (b) beam injection evaluated at 7 ms. Frame (c) illustrates the self-consistent electrostatic potential, $e \Phi/T_e$. }
 \label{Fig_1D2V_beam_ON_OFF}
\end{figure*}

Finally, we note significantly enhanced performance of the IMEX simulations using the coarse grid resolution. At the end of the run, the stable IMEX time step is $\SI{19}{\micro\second}$, the number of JFNK iterations is 65, and each time step requires approximately 1.8s on 512 cores. The entire \SI{20}{\milli\second} simulation completes in about 25 minutes. Although the numerical solution is not fully converged on the coarse grid, these fast, low-resolution runs are well suited for scoping studies and/or for providing informed initial conditions when searching for steady-state solutions.

\subsection{Simulations including a self-consistent E-field }

The results of the IMEX baseline-grid simulations including the effects of a self-consistent electric field [in Eq.~(\ref{adiabEl})] are shown in Fig.~\ref{Fig_1D2V_vparmu}(b) and Fig.~\ref{Fig_1D2V_lineouts}.  Compared to the corresponding results from simulations without an electric field, we observe a very similar solution in the high-energy beam region and slightly degraded confinement in the thermal background region. The latter can be attributed to the presence of electrostatic forces that repel ions from the trap. At the same time, the presence of density peaks at the turning points of high-energy (sloshing) ions leads to the formation of a potential well that can help confine low-energy ions. To illustrate this point, we compare the results of simulations with and without beam injection. Figure~\ref{Fig_1D2V_beam_ON_OFF} shows a zoom-in on the low-energy region, revealing enhanced confinement of low-energy ions. 

For the case without beam injection we observe the potential drop between the mirror center and the throat of $e \Delta \Phi /T_e \approx 7.6$, which is somewhat larger than the ambipolar Pastukhov potential inferred from the balance of electron and ion collisional losses from the trap, $\tau_{i,c}=\tau_{e,c}$. Specifically, estimating the ion confinement time as\cite{bing:1961, scheffel:2025} $\tau_{c,i} \sim 2.6 \log(R_m) \nu_i$, with the ion collision frequency $\nu_i$ given by Eq.~(\ref{nuLBD}), and using Eq.~(\ref{LossesAnalytic}) for the electron confinement time, $\tau_{c,e} = n_e | dn_e/dt |^{-1}$, we obtain a Pastukhov potential of $e\Phi_m/T_e = 6.3$. Accurately capturing the ambipolar potential, however, requires modeling kinetic electron effects, including their reflection from sheath regions at the device ends and unlike-species collisions. These processes are beyond the scope of the present simulations using the Boltzmann electron model. We further mention substantial limitations\cite{tyushev:2025} of the Boltzmann electron model in the expander region, i.e., outside the magnetic coil, where it predicts a potential drop $e \Delta \Phi \sim T_e \ln K$, which increases with the expansion ratio, $K$. Employing a kinetic electron response that captures cooling and temperature anisotropy effects, as well as electron reflection from the boundary sheath region, would resolve this issue and yield a potential solution~\cite{tyushev:2025} that quickly saturates for large values of $K$. We also note that the baseline grid resolution in velocity space is insufficient to adequately resolve the distribution function outside the trap, and the numerical results in that region are not fully converged. As discussed in Ref.~[\onlinecite{francisquez:2023}], this issue can be addressed, for instance, by introducing spatially varying normalization of the computational velocity space, which will be pursued in future work. 

Including the self-consistent electric field in Eq.~(\ref{adiabEl}) introduces collective phenomena, such as plasma sound waves, which are not captured by the "fixed-$E_z$" preconditioner model used in our simulations. These additional physical effects degrade the performance of the IMEX algorithms, with the most pronounced impact occurring at the beginning of the simulation when transient processes are strongest. To handle this initial stage, we employ an adaptive time-stepping algorithm from the PETSc library~\cite{petsc}, which adjusts the time step dynamically based on the solution behavior. As the simulation progresses and the system settles into a more quasi-steady state, the impact of the self-consistent electric field on solver performance diminishes. By the end of the run, we observe a modest slowdown of approximately 25\% when comparing baseline-grid $(256,128,192)$ simulations with and without electric fields (see Sec.~III~B), with a final stable time step of $\Delta t_\mathrm{fin}^{\text{CFL}} \approx \SI{4.7}{\micro\second}$. Specifically, about 70 JFNK iterations are required to reach the same solver tolerance of $10^{-4}$, and each step takes approximately 9.5s per step when run on 512 cores. This mild performance degradation is plausibly attributed to the condition $T_e \ll T_i$, as discussed in Sec.~III A. The stable time step for fully explicit time integration remains limited by advection in the $\vpll$-direction, $\Delta t_{\nabla B}^{\text{CFL}} \approx \SI{0.19}{\ns}$, with each step requiring approximately $\SI{1.0}{\s}$ similar to the case without electric fields. Thus, the IMEX approach provides a speedup of over $2.5\times10^3$ by the end of the run. It is also informative to compare this performance with recent explicit simulations of the WHAM system using the GKeyll code~\cite{francisquez:2023}. For the Boltzmann electron model and the original (non-smoothed) mirror force, GKeyll simulations employing the LBD collisional model require approximately 74 hours on 288 cores to simulate $\SI{72}{\micro\s}$ of physical time using the  $(N_z, N_{\vpll}, N_\mu)=(288, 64, 192)$ grid. Normalized by the number of grid cells and processor cores, this is roughly $1.8\times10^3$ times slower than the semi-implicit COGENT calculations employing the full nonlinear Fokker-Planck collision operator. 
  
\section{Conclusions}   

In this work, we report on the development and application of implicit-explicit (IMEX) time integration methods for continuum kinetic modeling of weakly collisional parallel plasma transport in magnetic mirror configurations. The IMEX approach, based on high-order additive Runge-Kutta methods and Jacobian-free Newton-Krylov solvers with modern algebraic multigrid preconditioning, enables stable and efficient simulations that overcome the severe time-step restrictions imposed by strong mirror forces in explicit schemes. As an illustrative example, we consider the problem of high-energy ion beam fueling of a weakly collisional mirror for parameters relevant to the WHAM facility. Both a simplified bounce-average model (BASM) for a square-shaped mirror and a fully kinetic model are employed to provide insights into ion species dynamics and to assess the performance of the numerical algorithms. 

The fully implicit BASM model is verified by reproducing analytic results for the Pastukhov problem and is used to examine the numerical convergence of the velocity-space discretization scheme. In particular, degraded convergence, limited to first-order accuracy, is observed when the transition layer between the confined and loss cone regions is insufficiently resolved. The BASM model also highlights the importance of an accurate ion-ion collision model by revealing significant differences in the numerical solution obtained using the full Fokker-Planck collision operator versus the simplified Lenard-Bernstein-Dougherty (LBD) operator often used in gyrokinetic simulations.

The fully kinetic 1D-2V model demonstrates the practical benefits of the IMEX approach, which involves an implicit treatment of the Vlasov operator. Time steps approximately $2.5 \times 10^4$ times larger than the CFL-limited explicit step are achieved, yielding about 2500x speedup compared to explicit methods while maintaining numerical stability. The current "fixed-$E_z$" preconditioner model does not account for collective oscillation processes such as plasma sound waves, which can limit its efficiency. Nevertheless, only modest performance degradation ($\sim 25 \%$ slowdown) is observed when self-consistent electric field effects -- modeled through quasi-neutrality and a Boltzmann electron response to the potential -- are included. This can be plausibly attributed to the relatively low electron temperature compared to the ion temperature in the parameter regime considered here, implying that corrections to the linear-response Jacobian from the self-consistent E-field model may be subdominant to the fixed-field ion advection terms. In addition, fifth-order upwind discretization of the Vlasov operator is found to be essential for minimizing numerical diffusion over long collisional transport timescales, whereas a third-order upwind scheme produces excessive numerical pollution.

The implicit technologies described in this work have the potential to enable comprehensive, integrated modeling of magnetic mirror systems using continuum gyrokinetic simulations. Our future work will focus on extending these methods to include radial nonuniformities, kinetic electrons, unlike-species collisions, and more complex magnetic geometries (e.g., tandem mirrors) relevant to next-generation mirror reactor designs. 
 
\begin{acknowledgments}
The authors are grateful to C. Forest, S. Frank, Y. Petrov, L. Ricketson, D. Martin and H. Johanses for fruitful discussions. This research was supported by the U.S. Department of Energy under contract DE-AC52-07NA27344. 
\end{acknowledgments}

\section*{data availability}
The data that support the findings of this study are available from the corresponding author upon reasonable request.

\bibliography{./Bibliography.bib}
\end{document}

%% file: definitions.tex
\newcommand{\vpll}{\mathrm{v}_{\parallel}}
\newcommand{\vpllhat}{\widehat{\mathrm{v}}_{\parallel}}

\newcommand{\vperp}{\mathrm{v}_{\perp}}